\documentclass[12pt]{article}
\usepackage[utf8]{inputenc}
\usepackage[T1]{fontenc}
\usepackage{graphicx}
\usepackage{authblk}
\usepackage{xurl}   %in case url needs linebreak
\usepackage{amsmath}
\usepackage{amssymb}
\usepackage{amsbsy}
\usepackage{hyperref}
\usepackage{comment}
\usepackage{subfig}
\usepackage{fancybox}
\usepackage{graphicx,color}
\usepackage{tcolorbox}
\usepackage{framed}
\usepackage{here} 
\usepackage{natbib}
\usepackage{geometry}
\usepackage{theorem}
\usepackage{booktabs}
\usepackage{marvosym}

\definecolor{darkblue}{rgb}{0,0,0.8}
\definecolor{darkgreen}{rgb}{0,0.8,0}

\definecolor{magenta}{rgb}{0.5,0,0.5}

\theoremstyle{break}
\theorembodyfont{}
\newtheorem{definition}{Definition}[section]

\definecolor{grau}{rgb}{0.5,0.5,0.5}

\definecolor{darkblue}{rgb}{0,0,0.35}
\usepackage{colortbl}

\title{A Bayesian Spatio-Temporal Model of Temperature- and Humidity-Related Mortality Using High-Resolution Climate Data}
\author[1]{Corinna Perchtold\footnote{\Letter \; {corinna.perchtold@jku.at}}}
\author[2]{Julia Eisenberg} \author[3]{Philipp Otto}

\affil[1]{Johannes Kepler University Linz
  }%
  \affil[2]{TU Wien}
\affil[3]{University of Glasgow}  
\date{\today}

\begin{document}
\maketitle

%\tableofcontents

\begin{abstract}\noindent
In this study, we introduce a novel and comprehensive extension of a Bayesian spatio-temporal disease mapping model that explicitly accounts for gender-specific effects of meteorological exposures. Leveraging fine-scale weekly mortality and high-resolution climate data from Austria (2002–2019), we assess how interactions between temperature, humidity, age, and gender influence mortality patterns.  
Our approach goes beyond conventional modelling by capturing complex dependencies through structured interactions across space–time, space–age, and age–time dimensions, allowing us to capture complex demographic and environmental dependencies.
The analysis identifies district-level mortality patterns and quantifies climate-related risks on a weekly basis, offering new insights for public health surveillance.

%In the present paper, we propose a novel extension of traditional Bayesian spatio-temporal models for disease mapping by incorporating a gender-specific component and relevant meteorological covariates. 
%We analyse the weekly spatio-temporal distribution of mortality in Austria over the period 2002–2019 and assess the extent to which this distribution can be explained by meteorological factors. Specifically, we look at the combination temperature/humidity, as well as at the demographic variables such as age, gender, and population size. Our approach is region-specific and incorporates interactions across space-time, space-age, and age-time dimensions. The analysis is based on the fine-grained observed mortality and weather data in Austria and seeks to identify patterns and correlations between mortality and age, temperature, and gender.

% We want to examine the weekly spatio-temporal distribution of mortalities in Austria over the time period $2002-2019$ and investigate to which extend this distribution can be explained by meteorological variables in relation to temperature and humidity but also by age, gender and population.
% Our modelling approach is region-specific, thus we introduce space-time, space-age, and age-time interactions.
% This analysis is based on real mortality and weather data from Austria and should detect coherences between mortality and age, temperature and gender.
\vspace{6pt}
	%\settowidth\labelwidth{{\it 2020 Mathematics Subject Classification: }}%
%	\par\noindent {\it 2020 Mathematics Subject Classification: }%
%	\rlap{Primary}\phantom{Secondary}
	%\newline\null\hskip\labelwidth
	%Secondary 
\end{abstract}

%as weekly maximum, minimum and mean temperature, weekly length of a dry spell, (lagged) heat and cold waves, weekly mean humidity and mean precipitation, the interaction between minimum or maximum temperature and humidity, but also by age, gender and population. Therefore, we set up a model for mortality rates that additionally includes space-time, space-age, and age-time interactions. % to gain knowledge about the spatial variation and temporal evolution of mortality.
%The main goal of this work is to detect coherences between mortality and age, temperature and gender. %In this paper, mortality rates by region, time, and age groups are provided to gain knowledge about the spatial variation and temporal evolution of mortality. The main goal of this work is to detect spatiotemporal patterns according to the different age groups.

%\section{Introduction}

%\textbf{PO: we should probably also cite the following papers: %\cite{khana2018bayesian},
%\cite{quick2017multivariate}, \cite{etxeberria2014evaluating}.}

\section{Introduction}

Understanding population mortality has long been central to demographic research, actuarial science, and public policy. In Austria, systematic recording began with the first modern census in 1869; historical mortality patterns were later reconstructed by Findl  (1979) and Gisser (1979)  \cite{findl, gisser}, while current period mortality tables are maintained by Statistics Austria, \cite{sa}. Traditional life tables, which summarise age- and sex-specific mortality rates, play a key role in (life) insurance, pension planning, and civil litigation \cite{booth}. However, such tables provide a static, annual snapshot and do not capture short-term fluctuations in mortality due to environmental or epidemiological stressors. In recent years, the availability of high-resolution data has enabled more refined analyses of mortality dynamics, especially in response to climatic exposures such as temperature and humidity.

% The first modern census in Austria was conducted in 1869. Life tables for the years prior to 1912 were later estimated by Findl (1979), while Gisser documented historical developments in births, deaths, and migration patterns \cite{findl, gisser}. For more recent data, modern period mortality tables are available from Statistics Austria: \url{https://www.statistik.at/en}.\medskip

% A period mortality table captures age- and sex-specific mortality rates for a given year. Such data play a crucial role in various areas, for instance, in civil law, where they are used to calculate financial settling following (i.e.\ car) accidents. One of the primary applications, however, is in (life) insurance, where life tables are essential for determining indemnity payments, annuities and the present value of future pensions -- topics that are vital to the financial security of our society, a review on the methods for mortality modelling and forecasting one finds, for instance in \cite{booth}. Without a doubt, life tables are among the most important tools for analysing demographic trends, which will significantly shape social development over the coming decades. \medskip

A clear illustration of how changes in mortality impact all dimensions of society is the steady increase in longevity. Over the last couple of decades, the term longevity has become increasingly prominent in research areas like medicine, actuarial sciences, finance, social security, among others, see for instance \cite{murphy} or a more recent work by \cite{scott}. Longevity means that people live a longer, however, not necessarily a healthier life. While longevity is undeniably positive on one hand, it has consistently raised concerns for governments and insurers. Indeed, combined with an ageing population, increased longevity places significant pressure on state pension systems and public healthcare services. In European Union, the old-age dependency ratio -- the number of individuals aged 65 or older per 100 people of working age -- is predicted to increase to 58.60\% by 2075, see \cite{oecd}, which would create a significant burden for societies if not mitigated quickly, see \cite{pitacco} on modelling longevity dynamics for pensions and annuity business. \medskip

However, alongside developments in longevity, medical and social science literature has increasingly highlighted temporary spikes in mortality assumed to be correlated to some specific weather events like heat, cold and/or humidity. To name just a few studies in this direction: \cite{eurowinter} addresses the cold exposure and winter mortality, \cite{gasparrini} look at the impact of the heat waves, \cite{basu} investigate epidemiological evidence,  \cite{diego} discuss the interplay between social-based reasons and extreme weather causing increased mortality in Madrid between 1917 and 1921. 
Similar to longevity, seasonal increases in mortality could impact every facet of people's lives. A spike in mortality is usually preceded by a hospitalisation or a medical treatment, causing apart from high costs, also staff shortages in ambulances and hospitals, see \cite{ambulance1}, \cite{ambulance2}.  Therefore, models that analyse the impact of age and gender on mortality in response to specific weather events are becoming increasingly important, this resembles the situation with COVID-19, see, for instance, Gleiss et al.\ \cite{covidaustria}.
\\A persistent increase in frequency of extreme weather events, causing mortality spikes, will call for actions raging from the reforms of the health system to different engineering/architectural solutions for big cities and rural areas, that could potentially be an ambulance desert. While Yiu et al.\ \cite{Torsten} do not address climate change, they examine the impact of deprivation on mortality rates. The development and importance of personalised treatment strategies within survival models is explored by Efthimiou et al.\ \cite{treatmentchoices}, with broader healthcare applications discussed by Gutzeit et al.\ \cite{healthcare}.

The described effect of long-term alterations in temperature and weather patterns is commonly referred to as climate change -- another term dominating the news. According to the European Environmental Agency ``Europe is the fastest warming continent in the world, and climate risks are threatening its energy and food security, ecosystems, infrastructure, water resources, financial stability, and people's health.'', see \url{https://www.eea.europa.eu/en/topics/in-depth/climate-change-impacts-risks-and-adaptation}.%the % breaks the url
\medskip

In the present paper, we examine the relationship between meteorological conditions and variations in mortality across Austria, using fine-grained, publicly available data. We propose a statistical mortality model tailored to estimate mortality rates at the district level.
% In the present paper, we investigate the correlation between the weather and the changes in mortality in Austria, using fine-grained, publicly available data. 
% We propose a statistical mortality model for Austria that specifically aims to determine mortality rates on a district level.
% \\Specifically we want to investigate how temperature affected mortality over the last $20$ years. The last Climate status reports  from Austria, \cite{KlimaAut20}, \cite{KlimaAut21}, \cite{KlimaAut22} and \cite{KlimaAut23}  all claim rising temperatures and dangers to human health that should not be underestimated. These include  the direct heat stress in summer (including low night-time cooling) and the changed spread of pathogens and their carriers.
% Moreover, \cite{Baldwin2023} highlights the need to investigate the influence of high levels of humidity and heat-related mortality.
We extend the age–space–time interaction model from \cite{Goicoa2016} and apply it to the real mortality data, in which  we consider age, gender, and high-resolution temperature and humidity data, all available on different temporal and spatial scales. Thus, the novelty in our framework is two-fold:% two important extensions to the framework of \cite{Goicoa2016} in the context of Austrian mortality data:
\begin{itemize}
    \item[(i)] we extend the statistical model by allowing explicitly for a gender dimension, essential for quantifying climate impacts on mortality, and
    \item[(ii)] it incorporates harmonised high-resolution meteorological and environmental covariates.
\end{itemize}
%We chose not to use Model 9 from the same paper, which adds a full age–space–time interaction, due to its high computational cost and limited improvement in performance as noted by the authors.
Adding gender is particularly important in our setting, as mortality patterns often differ between males and females -- especially in response to environmental stressors like extreme temperatures or humidity (see Figure \ref{fig:average_deaths}). Ignoring gender could obscure meaningful variation and lead to biased or overly smoothed results.
The necessity of dealing with the potential age and gender effects is clear, because the mortality rates do not stay constant over a person's lifetime and clearly depend on the gender, see for instance the life tables provided by Statistik Austria \cite{sa}. 
%We choose first order random walks for both time and age effects. 
Although embedding age-related effects in mortality rates is well-established in spatial and spatio-temporal disease mapping, it has only been partially analysed in the existing literature, see Goicoa et al.\ \cite{Goicoa2016} and references therein.
\\Specifically, this study aims to investigate the impact of temperature on mortality in Austria over the past two decades. Recent Austrian Climate Status Reports, see \cite{KlimaAut20}, \cite{KlimaAut21}, \cite{KlimaAut22}, consistently highlight rising temperatures and the associated risks to human health. These risks include direct heat stress during the summer months -- particularly due to insufficient night-time cooling -- as well as worsening of chronic health conditions such as cardiovascular, respiratory, diabetes, and mental health disorders.
%as shifts in the distribution of pathogens and their vectors. 
%Furthermore, 
%\\We build on the age–space–time interaction model from \cite{Goicoa2016} and apply it to the real mortality data.
The interaction terms are completely structured, and correspond to Type IV interactions as described
in Knorr-Held \cite{Knorr‐Held2000}.
Inference is done with INLA, \cite{Rue2009}, in combination with the PARDISO solver.\footnote{\url{www.panua.ch}} This approach is readily accessible in the free software R through the R-INLA package.\footnote{\url{https://www.r-inla.org/download-install}}
Similar work on spatio-temporal variations in mortality rates on subnational level with INLA is done, e.g., in \cite{khana2018bayesian}.

Baldwin et al.\ \cite{Baldwin2023} stress the importance of examining the combined effects of high humidity and heat on mortality. Indeed, as relative humidity increases, the body's ability to regulate temperature through perspiration diminishes, thereby exacerbating heat stress. Therefore, we examine whether the interaction between temperature and humidity has a statistically significant effect on mortality. To capture this interaction, we use the heat index -- a metric that reflects perceived temperature by incorporating both air temperature and relative humidity, see \cite{Steadman1979}.

%Our approach is novel in several ways. 
%We describe our model in the context of the Austrian mortality dataset.
%It builds upon Model $8$ from Goicoa et al.\ \cite{Goicoa2016}, extending it in two key ways: 
%\begin{itemize} 
%\item[(i)] by explicitly incorporating gender as a dimension, and 
%\item[(ii)] by including a set of meteorological and environmental covariates. 
%\end{itemize}
%We opted not to adopt Model $9$, which introduces a full age-space-time interaction term, due to its substantial computational cost and only marginal gains in model performance as reported by the authors of \cite{Goicoa2016} themselves.

%The inclusion of gender is essential in our context, as mortality patterns can differ significantly between males and females, see Figure \ref{fig:average_deaths}, - particularly in response to environmental factors such as temperature extremes or humidity. Ignoring this distinction could mask important variation and lead to biased or overly smoothed estimates of mortality rates.

The paper is organized as follows. Section 2 presents the data and details the selection of covariates. In Section 3, we introduce our mortality model and explain the INLA approach. Section 4 discusses the results, with a particular focus on the effects of temperature and humidity. Finally, Section 5 concludes the paper and outlines potential directions for future research.

\section{Data Description and Exploratory Data Analysis}
%%%%%%%%%%%%%%%%%%%%%%%%%%%%%%%%%%%%%%%%%%%%%%%%%

To assess the spatio-temporal distribution of mortality in Austria and its relation to meteorological stressors, we integrated multiple high-resolution datasets with demographic and environmental relevance. These include weekly mortality counts disaggregated by age, gender, and district, meteorological measurements from a dense network of monitoring stations, and derived covariates characterising temperature and humidity extremes. In the following subsections, we describe the data sources, the preprocessing steps applied to each component, and the methods used for spatial and temporal alignment, covariate engineering, and data integration. This structured approach ensures consistency across disparate data domains and allows for the robust estimation of gender-specific mortality patterns under varying climatic conditions.

\subsection{Mortality data and their spatial-temporal domain}

Our study is based on weekly mortality counts across Austria's $94$ administrative districts, covering the period from $2002$ to $2019$. The data were obtained from the Austrian Federal Statistical Office \cite{sa} and are disaggregated by:
%The weekly mortality data, detailing the number of deaths across Austria’s $94$ districts, was obtained from the Federal Institute of Statistics Austria, \cite{sa}. 
%The data is categorised by
\begin{itemize}
\item  four age groups: $0-64$, $65-74$, $75-84$, $85+$.
\item two genders: female and male,
\item for each calendar week from: $2002$ to $2019$.\footnote{
Due to data protection rules, more detailed age groups were not available. To keep weekly resolution, we used the given age groupings.}
%A more detailed breakdown of age groups was not possible due to privacy restrictions. Otherwise, a broader time interval than a calendar week would have been necessary.}
\end{itemize}

%The data is categorised by four age groups ($0-64$, $65-74$, $75-84$, $85+$) and two genders (male and female) for each calendar week from $2002$ to $2019$.

Corresponding gender-specific population data were retrieved from the STATcube database, providing annual population counts for each district, and age group. We assume population sizes remain constant within each calendar year.

%Corresponding gender-specific population data for these age groups and districts from $2002-2019$ was sourced from the STATcube statistical database. Since only one yearly population data point is available per district and age group,  we assume a constant population within each year.

%Using these datasets, we calculated mortality rates as the ratio of deaths to the population. As an example, Figure \ref{fig:mortality_rate_12} presents the mortality rate time series for every calendar week in $2012$.
\begin{figure}[t]
    \centering
\includegraphics[width=0.8\linewidth]{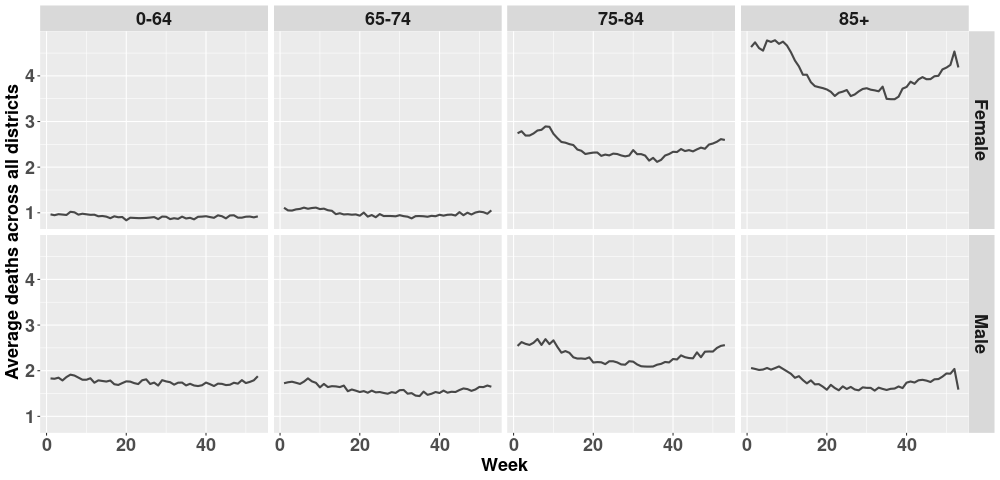}
    \caption{Average number of deaths across all districts from $2002-2019$.}
    \label{fig:average_deaths}
\end{figure}

Figure \ref{fig:average_deaths} presents the average weekly death counts across all districts, age groups, and genders. Consistent with demographic patterns, we observe higher mortality in males at younger ages and in females at older ages. Additionally, a seasonal pattern is evident, with pronounced winter peaks in the oldest age groups, underscoring the importance of environmental exposures in these vulnerable populations.

%We computed the average number of deaths for each age group and each gender across all districts, for the entire time horizon $2002-2019$, see Figure \ref{fig:average_deaths}. As one would expect, on average males have a higher number of deaths in the younger age groups, whereas females tend to die rather at an older age. 
%It can be seen that, for both males and females, the age groups with the highest mortality rates showcase an even higher rate in the winter months. Thus, a clear dependence on the weeks of a year can be seen for the most vulnerable groups of both genders.

%Next Figure \ref{fig:mortality-rate} illustrates the increased mortality rate of male compared to female. Therefore, we computed the mortality rate for each district, gender and age group, 

%\begin{figure}
%    \centering
   % \includegraphics[width=0.8\linewidth]{Average mortality rate all times.png}
    %\caption{Average male mortality rate in proportion to female one from $2002-2019$. }
    %\label{fig:mortality-rates}
%\end{figure}

%\begin{figure}
%    \centering
    %\includegraphics[width=\linewidth]{time series 2012 mortality rates.png}
    %\caption{Weekly mortality rate of females and males in 4 age groups for $2012$.}
    %\label{fig:mortality_rate_12}
%\end{figure}

%\begin{figure}
%    \centering
    %\includegraphics[width=\linewidth]{time series 2012 deaths.png}
    %\caption{Weekly number of deaths of females and males in 4 age groups for $2012$.}
    %\label{fig:death_rate_12}
%\end{figure}
\begin{figure}[t]
\centering
\includegraphics[width=\linewidth, trim=0 35mm 0 35mm,clip]{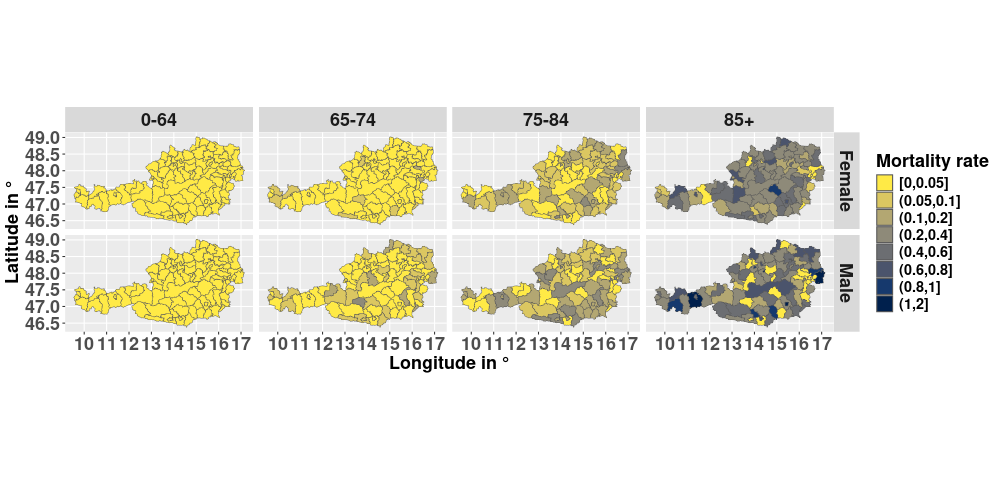}
    \caption{Spatial distribution of mortality rate of females and males in 4 age groups for $19.-25.3.2012$.}
    \label{fig:mortality_rate_w33}
\end{figure}

Spatial variability is substantial across both age and gender, as illustrated in Figure \ref{fig:mortality_rate_w33}, which shows district-level mortality rates for a selected week. The spatial heterogeneity observed motivates our Bayesian spatio-temporal approach, which explicitly accounts for regional and demographic structure.
The corresponding geographical data for Austria’s district boundaries and elevation were obtained from the Global Administrative Areas database.\footnote{\url{https://gadm.org/download_country.html}}

%For further analysis, we needed the geographical data of Austria's $94$ districts. We downloaded an elevation map of Austria with all borders of their subdivisions from the website of the Global Administrative Areas.\footnote{\url{https://gadm.org/download_country.html}}

%%%%%%%%%%%%%%%%%%%%%%%%%%%%%%%%%%%%%%%%%%%%%%%%%%%%%%

%\begin{itemize}
%    \item[2)] subgroups
%\end{itemize}

%%%%%%%%%%%%%%%%%%%%%%%%%%%%%%%%%%%%%%%%%%%%%%%%%%%%%
\subsection{Selection of meteorological data and its spatio-temporal resolution}

To investigate the potential environmental drivers of mortality, we collected meteorological data from the Austrian Central Institute for Meteorology and Geodynamics, known as GeoSphere. This dataset includes
%The Austrian Central Institute for Meteorology and Geodynamics, known as GeoSphere, provides a data hub for meteorological records collected across Austria and any period in the past. This study is based on
\begin{itemize}
    \item daily precipitation sums in mm,
    \item daily mean, minimum and maximum temperature data in $^\circ$C and
    \item daily mean level of humidity in $\%$ 
\end{itemize}
%daily precipitation sums (in mm) but also on daily mean, minimum and maximum temperature data (in $^\circ$C) and the daily mean level of humidity (in $\%$) 
recorded at $211$ monitoring stations across the country. These data were used to construct weekly summaries and define climate-related covariates.
%from all monitoring stations throughout Austria. 

% Approximately $2-2.3\%$ of values were missing (see Table \ref{fig:missing_data}). We addressed this using the \textit{mice} package in R, following the approach of \cite{Buuren2011}. Missing values were imputed for each station, scanning for gaps on a weekly basis.
%For each instance of missing data, we used all available observations from the same month to drew a sample value.

Approximately $2$–$2.3\%$ of the daily meteorological values were missing (see Table~\ref{fig:missing_data}). To impute these, we applied a simple stochastic imputation method: for each monitoring station, missing values were identified on a weekly basis, and each gap was filled by randomly sampling from the observed values at that station within the same calendar month. This preserves temporal structure and local climatic variation while avoiding parametric assumptions. The procedure was implemented using the \texttt{mice} package in \textsf{R}, which relies on a Markov Chain Monte Carlo (MCMC) algorithm to generate multiple imputed datasets \citep[see][for details]{Buuren2011}.

%To perform the imputation, we first divided the data by individual monitoring stations and systematically searched for missing values on a weekly basis. When missing values were identified, we gathered all available data for the corresponding month and drew a sample to estimate the missing meteorological variables.
%Due to an incomplete meteorological data set, Table \ref{fig:missing_data} shows the percentage of missing daily covariate data. We imputed the missing values with the \textit{mice} library in R, \cite{RTeam2011}. 
% For further information about this package, see \cite{Buuren2011}.
%To perform the imputation, we first divided the data by individual monitoring stations and systematically searched for missing values on a weekly basis. When missing values were identified, we gathered all available data for the corresponding month and drew a sample to estimate the missing meteorological variables.
 
\begin{table}[h]
%\fontsize{8pt}{8pt}{
\centering
\begin{tabular}{|l|r|r|r|r|r|r|}
\hline
\textbf{Temp mean} &\textbf{Temp min} & \textbf{Temp max} &  \textbf{Humidity mean} & \textbf{Precip sum}
\\
\hline
$2.07\%$ & $2.07\%$ & $2.07\%$  &$2.33\%$   &$2.33\%$ 
\\
\hline
\end{tabular}
    \caption{Percentage of missing daily covariate data.}
\label{fig:missing_data}
\end{table}

\subsection{Feature engineering and covariate selection}

% To reflect both typical and extreme weather conditions, we derived a set of weekly covariates: % from the imputed meteorological data set:
A critical step in modelling climate-related mortality involves constructing covariates that accurately capture both typical and extreme environmental exposures. To this end, we derived a set of weekly meteorological covariates that represent central tendencies as well as short-term anomalies in temperature and humidity conditions:
\begin{itemize}
    \item (Calendar) Weekly means of humidity, minimum, and maximum temperatures.
    \item Indicator variables for hot and cold weeks, based on definitions by AGES, the Austrian Agency for Health and Food Safety.\footnote{\url{https://www.ages.at/umwelt/klima/klimawandelanpassung/hitze}}
\item Derived variables reflecting lagged effects of extreme weather (e.g., “last week was dry”), adapted from \cite{Han2017}.
\item 
Classification of weeks using heat index levels (based on Steadman \cite{Steadman1979}) to identify periods of elevated temperature-humidity related risk.
\end{itemize}

%From the imputed meteorological data set, we calculated the weekly (calendar) mean humidity, as well as the average minimum and maximum temperatures, for each of the $211$ monitoring stations.
%\\On top, we introduced new covariates to account for hot and cold weeks. These are based on the definitions by AGES, the Austrian Agency for Health and Food Safety.\footnote{\url{https://www.ages.at/umwelt/klima/klimawandelanpassung/hitze}}

\begin{definition}
\begin{enumerate}
\item 
A \textit{hot week} is defined as at least three consecutive days with daily minimum temperature above $18 ^\circ$C, while a \textit{cold week} is understood as at least three consecutive days with daily minimum temperature below $0 ^\circ$C. 
To examine potential lagged effects of heat or cold waves on mortality, we introduced a covariate called \textit{Last week was dry}, following the approach in \cite{Han2017}. 
\item A (hot or cold) \textit{dry period} is described as a period of at least $3$ consecutive dry days (i.e.\ no precipitation) with daily mean temperature above the $95$th percentile or below the $5$th percentile. In our case, the $95$th percentile corresponds to daily mean temperature above $21.8^\circ$C and daily mean temperature below the $5$th percentile to $-5^\circ$C.
\item If a given week has met the criteria from 2., the following week is labelled as \textit{Last week was dry}, indicating whether mortality rates may increase as a delayed response to extreme heat or cold events.
%Further we implemented a heat wave according to \cite{Kysely2000}, which is defined as at least two successive days during which the maximum temperature exceeds $29^\circ$C. We labeled such a week \textit{Tropical week}.
\item \textit{Super cold weeks} are defined as weeks containing at least two consecutive days with minimum temperature below $-5^\circ$C.
\item To account for mild periods, we introduce the covariate \textit{Mild week}, which is characterised by at least three consecutive days with a daily mean temperature above $2^\circ$C and below $9^\circ$C.
\item Given Austria’s diverse topography—where two-thirds of the landscape is mountainous, with elevations ranging from $112$ to $3750$ meters above sea level -- we also included \textit{Elevation} as a covariate.
\end{enumerate} 
\end{definition}
%\begin{figure}[t]
%    \centering
%    \includegraphics[scale=1.5]{index.jpg}
%    \caption{Heat index calculated according to \cite{Steadman1979}. Temperature values (vertical) and humidity values (horizontal). Source: \cite{Gerics}}
%    \label{fig:heat-index}
%\end{figure}

\begin{definition}
Additionally, using the calendar weekly mean humidity data, we implemented four heat-index categories, based on Steadman \cite{Steadman1979}. With the heat-index levels $3$ to $6$ in Figure \ref{fig:heat-index} we define
\begin{itemize}
 \item \textit{Strong discomfort}: index 3 (pink area),
 \item \textit{Severe malaise}: index 4 (yellow area), 
 \item \textit{Increased risk}: index 5 (orange area),
 \item \textit{Serious risk}: index 6 (red area). 
 \end{itemize}
A week was classified under one of these categories \textit{if at least one day met the respective criteria}.

\end{definition}
\begin{table}[htbp]
\centering
\resizebox{\textwidth}{!}{%
\begin{tabular}{c|*{16}{c}}
\toprule
 & 25\% & 30\% & 35\% & 40\% & 45\% & 50\% & 55\% & 60\% & 65\% & 70\% & 75\% & 80\% & 85\% & 90\% & 95\% & 100\% \\
 \hline
\midrule
42 &\cellcolor{orange} 48 & \cellcolor{orange} 50 & \cellcolor{orange} 52 & \cellcolor{red}55 & \cellcolor{red} 57 & \cellcolor{red} 59 & \cellcolor{red} 62 & \cellcolor{red} 64 & \cellcolor{red} 66 & \cellcolor{red} 68 & \cellcolor{red} 71 & \cellcolor{red} 73 & \cellcolor{red} 75 & \cellcolor{red} 77 & \cellcolor{red} 80 & \cellcolor{red} 82 \\
41 &\cellcolor{orange} 46 & \cellcolor{orange} 48 & \cellcolor{orange} 51 & \cellcolor{orange} 53 &\cellcolor{red} 55 & \cellcolor{red} 57 & \cellcolor{red} 59 & \cellcolor{red} 61 & \cellcolor{red} 64 & \cellcolor{red} 66 & \cellcolor{red} 68 & \cellcolor{red} 70 & \cellcolor{red} 72 & \cellcolor{red} 74 & \cellcolor{red} 76 & \cellcolor{red} 79 \\
40 &\cellcolor{yellow} 45 & \cellcolor{orange} 47 & \cellcolor{orange} 49 & \cellcolor{orange} 51 & \cellcolor{orange} 53 & \cellcolor{red}55 & \cellcolor{red} 57 & \cellcolor{red} 59 & \cellcolor{red} 61 & \cellcolor{red} 63 & \cellcolor{red} 65 & \cellcolor{red} 67 & \cellcolor{red} 69 & \cellcolor{red} 71 & \cellcolor{red} 73 & \cellcolor{red}75 \\
39 & \cellcolor{yellow} 43 & \cellcolor{yellow} 45 &\cellcolor{orange} 47 & \cellcolor{orange} 49 & \cellcolor{orange} 51 & \cellcolor{orange} 53 &\cellcolor{red} 55 & \cellcolor{red} 57 & \cellcolor{red} 59 & \cellcolor{red} 61 & \cellcolor{red} 63 & \cellcolor{red} 65 & \cellcolor{red} 66 & \cellcolor{red} 68 & \cellcolor{red} 70 & \cellcolor{red} 72 \\
38 &\cellcolor{yellow} 42 & \cellcolor{yellow}44 & \cellcolor{yellow}45 &\cellcolor{orange} 47 & \cellcolor{orange}49 & \cellcolor{orange}51 & \cellcolor{orange}53 &\cellcolor{red} 55 & \cellcolor{red} 56 & \cellcolor{red} 58 & \cellcolor{red} 60 & \cellcolor{red} 62 & \cellcolor{red} 64 & \cellcolor{red}66 & \cellcolor{red} 67 & \cellcolor{red} 69 \\
37 &\cellcolor{yellow} 40 & \cellcolor{yellow}42 & \cellcolor{yellow}44 & \cellcolor{yellow}45 &\cellcolor{orange} 47 & \cellcolor{orange}49 & \cellcolor{orange}51 & \cellcolor{orange} 52 &\cellcolor{red} 54 & \cellcolor{red} 56 & \cellcolor{red} 58 & \cellcolor{red} 59 & \cellcolor{red} 61 & \cellcolor{red} 63 &\cellcolor{red} 65 & \cellcolor{red} 66 \\
36 &\cellcolor{pink} 39 &\cellcolor{yellow} 40 & \cellcolor{yellow} 42 & \cellcolor{yellow} 44 & \cellcolor{yellow} 45 & \cellcolor{orange}47 & \cellcolor{orange} 49 & \cellcolor{orange} 50 & \cellcolor{orange} 52 &\cellcolor{red} 54 & \cellcolor{red} 55 & \cellcolor{red} 57 & \cellcolor{red} 59 &\cellcolor{red} 60 & \cellcolor{red} 62 & \cellcolor{red} 63 \\
35 &\cellcolor{pink} 37 &\cellcolor{pink}39 &\cellcolor{yellow} 40 & \cellcolor{yellow}42 & \cellcolor{yellow}44 & \cellcolor{yellow}45 &\cellcolor{orange} 47 &\cellcolor{orange} 48 & \cellcolor{orange}50 & \cellcolor{orange}51 & \cellcolor{orange}53 & \cellcolor{red}54 & \cellcolor{red} 56 &\cellcolor{red} 58 & \cellcolor{red} 59 & \cellcolor{red} 61 \\
34 & \cellcolor{pink}36 & \cellcolor{pink}37 & \cellcolor{pink} 39 &\cellcolor{yellow} 40 & \cellcolor{yellow}42 & \cellcolor{yellow}43 & \cellcolor{yellow}45 &\cellcolor{orange} 46 & \cellcolor{orange}48 & \cellcolor{orange}49 & \cellcolor{orange}51 & \cellcolor{orange}52 &\cellcolor{red} 54 & \cellcolor{red} 55 & \cellcolor{red} 57 & \cellcolor{red} 58 \\
33 & 34 & \cellcolor{pink}36 &\cellcolor{pink} 37 & \cellcolor{pink}39 &\cellcolor{yellow} 40 & \cellcolor{yellow}41 & \cellcolor{yellow}43 & \cellcolor{yellow}44 &\cellcolor{orange} 46 & \cellcolor{orange}47 & \cellcolor{orange}48 & \cellcolor{orange}50 & \cellcolor{orange}51 & \cellcolor{orange}53 &\cellcolor{red} 54 & \cellcolor{red} 55 \\
32 & 33 & 34 &\cellcolor{pink} 36 & \cellcolor{pink} 37 & \cellcolor{pink} 38 & \cellcolor{yellow}40 & \cellcolor{yellow}41 & \cellcolor{yellow}42 & \cellcolor{yellow}44 & \cellcolor{yellow} 45 &\cellcolor{orange} 46 & \cellcolor{orange}48 & \cellcolor{orange}49 & \cellcolor{orange}50 & \cellcolor{orange}52 & \cellcolor{orange} 53 \\
31 & 32 & 33 & 34 &\cellcolor{pink} 35 & \cellcolor{pink} 37 & \cellcolor{pink} 38 &\cellcolor{yellow} 39 & \cellcolor{yellow} 40 & \cellcolor{yellow}42 & \cellcolor{yellow}43 & \cellcolor{yellow}44 & \cellcolor{yellow}45 &\cellcolor{orange} 47 & \cellcolor{orange}48 & \cellcolor{orange}49 & \cellcolor{orange}50 \\
30 & 30 & 32 & 33 & 34 &\cellcolor{pink} 35 & \cellcolor{pink}36 & \cellcolor{pink}37 & \cellcolor{pink}39 &\cellcolor{yellow} 40 & \cellcolor{yellow}41 & \cellcolor{yellow}42 & \cellcolor{yellow}43 & \cellcolor{yellow}45 &\cellcolor{orange} 46 & \cellcolor{orange}47 & \cellcolor{orange}48 \\
29 & 29& 30 & 31 & 32 & 33 &\cellcolor{pink} 35 & \cellcolor{pink}36 & \cellcolor{pink}37 & \cellcolor{pink}38 &\cellcolor{pink}39 &\cellcolor{yellow} 40 & \cellcolor{yellow}41 & \cellcolor{yellow}42 & \cellcolor{yellow}43 & \cellcolor{yellow}45 &\cellcolor{orange} 46  \\
28 & 28 & 29 & 30 & 31 & 32 & 33 & 34 &\cellcolor{pink} 35 & \cellcolor{pink} 36 & \cellcolor{pink} 37 & \cellcolor{pink} 38 & \cellcolor{pink} 39 &\cellcolor{yellow} 40 &\cellcolor{yellow} 41 & \cellcolor{yellow}42 & \cellcolor{yellow}43  \\
27 & 27 & 27 & 28 & 29 & 30 & 31 & 32 & 33 & 34 &\cellcolor{pink} 35 & \cellcolor{pink}36 & \cellcolor{pink}37 & \cellcolor{pink}38 & \cellcolor{pink}39 & \cellcolor{yellow}40 &\cellcolor{yellow} 41 \\
26 & 26 & 26 & 27 & 28 & 29 & 30 & 31 & 32 & 33 & 34 & 34 &\cellcolor{pink} 35 & \cellcolor{pink}36 & \cellcolor{pink}37 & \cellcolor{pink}38 & \cellcolor{pink}39 \\
25 & 25 & 25 & 26 & 27 & 27 & 28 & 29 & 30 & 31 & 32 & 33 & 34 & 34 &\cellcolor{pink} 35 & \cellcolor{pink}36 & \cellcolor{pink}37 \\
24 & 24 & 24 & 24 & 25 & 26 & 27 & 28 & 28 & 29 & 30 & 31 & 32 & 33 & 33 & 34 &\cellcolor{pink} 35\\
23 & 23 & 23 & 23 & 24 & 25 & 25 & 26 & 27 & 28 & 28 & 29 & 30 & 31 & 32 & 32 & 33 \\
22 & 22 & 22 & 22 & 22 & 23 & 24 & 25 & 25 & 26 & 27 & 27 & 28 & 29 & 30 & 30 & 31 \\
\bottomrule
\end{tabular}%
}
\caption{Heat index calculated according to \cite{Steadman1979}. Temperature values (vertical) and humidity values (horizontal).} %Source: \cite{Gerics}}
    \label{fig:heat-index}
\end{table}
%%%%%%%%%%%%%
\begin{figure}
     \centering
     \includegraphics[width=\textwidth, trim=0 25mm 0 20,clip]{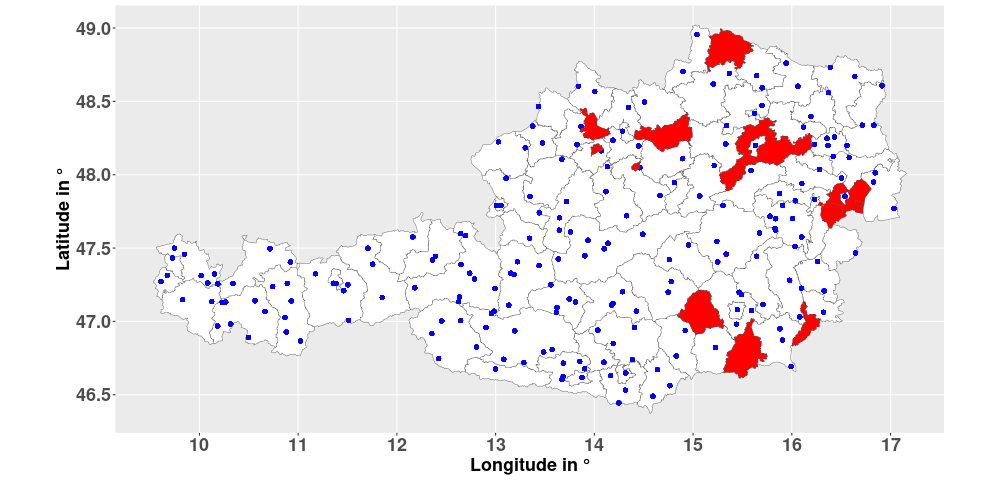}
     \caption{Red-coloured districts are the ones without blue monitoring stations.}
     \label{fig:districts}
 \end{figure}
\subsection{Data fusion and spatial alignment of meteorological and mortality data}

% To merge meteorological data with district-level mortality data, we spatially aggregated point-based weather data to the district level. Specifically, we averaged all monitoring station values within each federal state and assigned these values to the corresponding districts. Districts without any monitoring stations (12 out of 94, see Figure \ref{fig:missing_data}) have no covariate values and were handled using $k$-nearest neighbour imputation ($k=3$), based on Euclidean distance to nearby stations \cite{Fix89,Cover1967}.

A key challenge in our analysis lies in integrating meteorological observations, which are recorded at irregularly spaced monitoring stations, see Figure \ref{fig:districts}, with mortality data aggregated at the administrative district level. To address this, we first spatially aggregated point-based weather data by averaging the weekly covariate values across all stations within each federal state, subsequently assigning these values to the corresponding districts. However, 12 out of 94 districts lacked direct station coverage (see Figure~\ref{fig:missing_data}), resulting in missing environmental information. For these districts, we employed a spatial interpolation method based on $k$-nearest neighbour imputation ($k=3$), using Euclidean distances to nearby stations to assign surrogate values \cite{Cover1967,Fix89}. This fusion of high-resolution meteorological data with coarser, region-based mortality records enables the incorporation of environmental exposures at relevant spatial and temporal scales, a critical step in identifying localised vulnerability patterns in public health surveillance.

%Next, we merge the meteorological data with the mortality data.
%For this, we adopt a straightforward approach by aggregating point-based meteorological data into district-level data. Specifically, we average the calendar weekly covariate values across all monitoring stations within each federal state, thereby obtaining weekly district-level values.

%\begin{itemize}
%    \item[\textbf{2.5}]  \textbf{Imputation of missing values}
%\end{itemize}

%In total, $12$ out of $94$ districts have no monitoring stations, see Figure \ref{fig:districts}, and hence no covariate values. Therefore, we had to use the $k$-nearest neighbour (kNN) algorithm,  with $k=3$, % for regression in order to interpolate these values. For more information on kNN, have a look at \cite{Cover1967} and  \cite{Fix89}. %Briefly spoken, the kNN algorithm determines $k$ measurement stations, which in our case are closest to the missing district  to be predicted by the Euclidean distance measure, and then uses the average of the $k$ values and assigns it to the target location. 

% \subsubsection{Final choice of covariates}
%Before proceeding with the inference process, we wanted to select those covariates that best explained the mortality data. Therefore, we ran a

\section{Space-Time Modelling and Estimation}

Building on the integrated dataset described in the previous section, we now introduce the spatio-temporal modelling framework used to analyse gender-specific mortality patterns. We propose a gender-specific extension of a spatio-temporal model for mortality rates in Austria. The model includes $13$ fixed effects and random effects for space, time, and age to capture region-specific patterns in the data, see Section \ref{sec:mortality}.
The resulting model is then rewritten in a Bayesian hierarchical form and fitted using the INLA framework as described in Section \ref{sec:Bayesian_model}.

\subsection{Mortality model} \label{sec:mortality}

%\color{blue}
%We describe our model in the context of the Austrian mortality dataset.
%Our model builds upon Model $8$ from Goicoa et al.\ \cite{Goicoa2016}, extending it in two key ways: 
%\begin{itemize} 
%\item[(i)] by explicitly incorporating gender as a dimension, and 
%\item[(ii)] by including a set of meteorological and environmental covariates. 
%\end{itemize}
%We opted not to adopt Model $9$, which introduces a full age-space-time interaction term, due to its substantial computational cost and only marginal gains in model performance as reported by the authors of \cite{Goicoa2016} themselves.

%The inclusion of gender is essential in our context, as mortality patterns can differ significantly between males and females, see Figure \ref{fig:average_deaths}, - particularly in response to environmental factors such as temperature extremes or humidity. Ignoring this distinction could mask important variation and lead to biased or overly smoothed estimates of mortality rates.\\
%\color{black}

Let $y_{ijt,k}$, $n_{ijt,k}$, and $r_{ijt,k}$ be the number of deaths, the population at risk, and the mortality rate, respectively, in district $i$ ($i = 1,\ldots, 94$) and over the weeks of the years $2002,\ldots, 2019$, i.e.,  $t$ ($t =1,\ldots, 939$), for different age group $j$ ($j = 1,\ldots,4$) and gender $k$ ($k=1,2$).  

Then, conditional on the rate $r_{i jt,k}$, the number of deaths $y_{i jt,k}$ is assumed to follow a Poisson distribution
\begin{align*}
   & y_{i jt,k} \mid r_{i jt,k} \sim \text { Poi}\left(\mu_{i jt,k}=n_{i jt,k} r_{i jt,k}\right),\\\notag
    &\log(\mu_{i jt,k})=\log( n_{i jt,k})+\log( r_{i jt,k}),
\end{align*} 
where $\log( n_{i jt,k})$ denotes an offset.

%Since the mean $\mu_{i jkt}$ is not necessarily linearly related to the data, it is linked to a structured additive predictor $\eta_{ijkt}$ through a link function $g(\cdot)$, i.e., $g(\mu_{ijkt})=\eta_{ijkt}$. Due to the Poisson distribution, we have $g(\cdot)=\log(\cdot)$ and $
%\eta_{ijkt}=\log n_{i jk  t}+\log r_{i j kt}$
%where the logarithm of the population at risk, $\log( n_{ijkt})$, denotes an offset. 

The model is defined in terms of the mortality rate $\log(r_{i jt,k})$, incorporating both covariates to capture potential risk factors and random effects to account for unobserved sources of variability. As fixed effects at the district-level, we include the calendar weekly covariates described earlier.

%We define the model further by describing %Next, we set up the model for the mortality rate $\log(r_{ijkt})$. We include several covariates to quantify risk factors and random effects to deal with other sources of variability. We choose the following fixed effects on district level: the aforementioned calendar weekly covariates.
%Also included the elevation, since we want to investigate if temperature exposure and mortalities are linked to lowlands or rather mountainous regions.
To account for district-specific dependencies in space, time and age we %include the following random effects:
represent the underlying spatial district pattern by $\phi_{i,k}$, a normally distributed random effect with Leroux prior as proposed in \cite{Leroux2000}, i.e., $\pmb{\phi_k}=(\phi_{1,k},\ldots,\phi_{94,k})^T$. % with
%\textcolor{blue}{@Philipp: passt das so, dass ich bei den $\Sigma$'s auch den $k$-Parameter eingefügt habe?}
The terms $\delta_{j,k}$ and $\psi_{t,k}$ are normally distributed age and temporal random effects, respectively. Both are modelled through a random walk of order $1$ (RW$1$) over the $4$ different age groups,  $0-64$, $65-74$, $75-84$ and $85+$, and the $939$ weeks of the years $2002$-$2023$, i.e.,
$\pmb{\delta_k}=(\delta_{1,k},\ldots, \delta_{4,k})^T$ with 
\begin{align*}
    \pmb{\phi_k}&\sim N(0, \sigma^2_{\pmb{\phi_k}}\Sigma_{\pmb{\phi_k}}) \text{ and }    \Sigma_{\pmb{\phi_k}}=(\lambda_{\pmb{\phi_k}} R_{\pmb{\phi_k}}+(1-\lambda_{\pmb{\phi_k}})I_{\pmb{\phi_k}})^{-1}.\\\notag
        \pmb{\delta_k}&\sim N(0,\sigma^2_{\pmb{\delta_k}} \Sigma_{\pmb{\delta_k}}) \text{ and } \Sigma_{\pmb{\delta_k}}=R^{-}_{\pmb{\delta_k}},\\\notag
    \pmb{\psi_k} &\sim N(0, \sigma^2_{\pmb{\psi_k}} \Sigma_{\pmb{\psi_k}}) \text{ and } \Sigma_{\pmb{\psi_k}}=R_{\pmb{\psi_k}}^{-}.
\end{align*}
The matrix $R_{\pmb{\phi_k}}$ is a structure matrix that can be interpreted as the spatial neighbourhood matrix, $I_{\pmb{\phi_k}}$ is an identity matrix of dimension $94\times 94$. The parameter $\lambda_{\pmb{\phi_k}} \in [0, 1]$ defines the degree of the spatial dependency. For more information on the Leroux prior read \cite{Ugarte2014}.
%The terms $\delta_{j,k}$ and $\psi_{t,k}$ are normally distributed age and temporal random effects, respectively. Both are modeled through a random walk of order $1$ (RW$1$) over the $4$ different age groups,  $0-64$, $65-74$, $75-84$ and $85+$, and the $939$ weeks of the years $2002$-$2023$, i.e., $\pmb{\delta_k}=(\delta_{1,k},\ldots, \delta_{4,k})^T$ and $\pmb{\psi_k}=(\psi_{1,k},\ldots,\psi_{939,k})^T$ with 
%\begin{align*}
%    \pmb{\delta_k}&\sim N(0,\sigma^2_{\pmb{\delta_k}} \Sigma_{\pmb{\delta_k}}) \text{ and } \Sigma_{\pmb{\delta_k}}=R^{-}_{\pmb{\delta_k}},\\
%    \pmb{\psi_k} &\sim N(0, \sigma^2_{\pmb{\psi_k}} \Sigma_{\pmb{\psi_k}}) \text{ and } \Sigma_{\pmb{\psi_k}}=R_{\pmb{\psi_k}}^{-}.
%\end{align*}
The matrices $R_{\pmb{\delta_k}}$ and $R_{\pmb{\psi_k}}$ are known structure matrices, i.e., temporal neighbourhood matrices, corresponding to first order random walks for age and time, respectively, and can be looked up in \cite[Page 95]{Rue2005}. The symbol $^{-}$ represents the Moore-Penrose generalized inverse of a matrix, since these matrices are non-negative definite.

We also include space–age, space–time, and time-age interaction effects, denoted by $\zeta_{i, j,k}^1$, $\zeta_{i, t,k}^2$ and $\zeta_{j, t,k}^3$, respectively. We assume, that these vectors $\pmb{\zeta^1_k}=(\zeta^1_{1,1,k},\ldots, \zeta^1_{94,4,k})^T$, $\pmb{\zeta^2_k}=(\zeta^2_{1,1,k},\ldots, \zeta^2_{94,939,k})^T$ and $\pmb{\zeta^3_k}=(\zeta^3_{1,1,k},\ldots, \zeta^3_{4,939,k})^T$ are normally distributed with separable covariance matrices given by the Kronecker product of the marginal covariances, i.e.,
\begin{align*}
\pmb{\zeta^1_k}&\sim N(0, \sigma^2_{\pmb{\zeta^1_k}} \Sigma_{\pmb{\zeta^1_k}}) \text{ and } \Sigma_{\pmb{\zeta^1_k}}= \Sigma_{\pmb{\phi_k}} \otimes \Sigma_{\pmb{\delta_k}}, \\\notag
\pmb{\zeta^2_k}&\sim N(0, \sigma^2_{\pmb{\zeta^2_k}} \Sigma_{\pmb{\zeta^2_k}}) \text{ and } \Sigma_{\pmb{\zeta^2_k}}=\Sigma_{\pmb{\phi_k}} \otimes \Sigma_{\pmb{\psi_k}},  \\\notag
\pmb{\zeta^3_k}&\sim N(0, \sigma^2_{\pmb{\zeta^3_k}} \Sigma_{\pmb{\zeta^3_k}}) \text{ and } \Sigma_{\pmb{\zeta^3_k}}=\Sigma_{\pmb{\psi_k}} \otimes \Sigma_{\pmb{\delta_k}}.\notag
\end{align*}
These effects should capture the specificities of the districts in each time point and for each age group. 
Therefore, we have opted for interaction terms $\pmb{\zeta^1_k}$, $\pmb{\zeta^2_k}$, $\pmb{\zeta^3_k}$ of Type IV, described by \cite{Knorr‐Held2000}, which detail the evolution of mortality rates with age but the interactions also accounts for similar temporal trends in neighbouring regions, as well as in contiguous age groups. 
It should be noted that the structure matrices presented above suffer from rank deficiency problems. 
Therefore, specific constraints have to be added to guarantee identifiability of these interaction terms  and to avoid confounding with the main effects, see  \cite{Goicoa2018} or \cite{Schroedle2011}. % a comprehensive list for all four possible interaction terms is provided for different interaction terms.
Altogether the model for the mortality rates on district level looks like
\begin{align}\label{eq:mortality_rates_equation}
\log( r_{i j t,k})&=\alpha_k+ \gamma_{1,k} \text{ scale(Temp max mean)}+\gamma_{2,k} \text{ scale(Temp min mean)}\\
&+ \gamma_{3,k} \text{ scale(Humidity mean)}\notag\\
&+\gamma_{4,k} \text{ Strong discomfort}+\gamma_{5,k} \text{ Severe malaise}\notag\\
&+\gamma_{6,k} \text{ Increased risk}+\gamma_{7,k} \text{ Serious risk}\notag\\
&+\gamma_{8,k}\text{ Mild week}+  \gamma_{9,k}\text{ Hot week}\notag\\
&+\gamma_{10,k}  \text{ Cold Week}+\gamma_{11,k} \text{ Super cold week}\notag\\
&+\gamma_{12,k}\text{ Last week was dry}+ \gamma_{13,k}\text{ Elevation}\notag\\
&+\phi_{i,k}+\delta_{j,k}+\psi_{t,k}+\zeta_{i, j,k}^1+\zeta_{i, t,k}^2+\zeta_{j, t,k}^3,\notag
\end{align}
with intercept $\alpha_k$ quantifying the logarithm of the global risk for each gender.  %The coefficients of the covariates $\gamma_1,\ldots, \gamma_{14}$ to be estimated.

\subsection{Bayesian hierarchical model}\label{sec:Bayesian_model}

The way we have constructed our model allows to place it in the framework of a Bayesian hierarchical model with three levels:
\begin{itemize}
    \item the gender-specific data model $\pi(\pmb{y_k}|\pmb{x_k})$ with the observations $\pmb{y_k}$ conditioned on the latent effects $\pmb{x_k}=\{\alpha_k, \gamma_{1,k},
\ldots, \gamma_{13,k}, \pmb{\phi_k}, \pmb{\delta_k}, \pmb{\psi_k}, \pmb{\zeta^1_k}, \pmb{\zeta^2_k}, \pmb{\zeta^3_k}   \}$,
\item the latent Gaussian model $\pi(\pmb{x_k}|\pmb{\theta_k})$,
\item and  %the distribution of 
the hyperparameters $\pi(\pmb{\theta_k})$ with $\pmb{\theta_k}=\{\tau_{\pmb{\phi_k }}, \lambda_{\pmb{\phi_k}},
\tau_{\pmb{\delta_k}},\tau_{\pmb{ \psi_k}} , \tau_{\pmb{\zeta^1_k } }, \tau_{\pmb{ \zeta^2_k} },
\tau_{\pmb{ \zeta^3_k} } \}$ given by the precision parameters $\tau_{\pmb{\phi_k }}=1/\sigma^2_{\pmb{\phi_k }}$, $\tau_{\pmb{\delta_k}}=1/\sigma^2_{\pmb{ \delta_k} }$, $\tau_{\pmb{ \psi_k}}=1/\sigma^2_{\pmb{ \psi_k}}$, $\tau_{\pmb{\zeta^1_k } }=1/\sigma^2_{\pmb{\zeta^1_k } }$, $\tau_{\pmb{ \zeta^2_k} }=1/\sigma^2_{\pmb{ \zeta^2_k} }$ and $\tau_{\pmb{ \zeta^3_k} }=1/\sigma^2_{\pmb{ \zeta^3_k} }$. 
\end{itemize}
%\textcolor{blue}{@Philipp: auch hier wurde k-Abhängigkeit eingefügt. Passt das so?}

%Firstly, the \textcolor{blue}{gender-specific} data model $\pi(\pmb{y_k}|\pmb{x_k})$ with the observations $\pmb{y_k}$ conditioned on the latent effects $\pmb{x_k}=\{\alpha_k, \gamma_{1,k},
%\ldots, \gamma_{13,k}, \pmb{\phi_k}, \pmb{\delta_k}, \pmb{\psi_k}, \pmb{\zeta^1_k}, \pmb{\zeta^2_k}, \pmb{\zeta^3_k}   \}$. 
%Secondly, the latent Gaussian model $\pi(\pmb{x_k}|\pmb{\theta_k})$. % with precision matrix $Q$. 
%Thirdly, the distribution of the hyperparameters $\pi(\pmb{\theta_k})$ with $\pmb{\theta_k}=\{\tau_{\pmb{\phi_k }}, \lambda_{\pmb{\phi_k}},
%\tau_{\pmb{\delta_k}},\tau_{\pmb{ \psi_k}} , \tau_{\pmb{\zeta^1_k } }, \tau_{\pmb{ \zeta^2_k} },
%\tau_{\pmb{ \zeta^3_k} } \}$ given by the precision parameters $\tau_{\pmb{\phi_k }}=1/\sigma^2_{\pmb{\phi_k }}$, $\tau_{\pmb{\delta_k}}=1/\sigma^2_{\pmb{ \delta_k} }$, $\tau_{\pmb{ \psi_k}}=1/\sigma^2_{\pmb{ \psi_k}}$, $\tau_{\pmb{\zeta^1_k } }=1/\sigma^2_{\pmb{\zeta^1_k } }$, $\tau_{\pmb{ \zeta^2_k} }=1/\sigma^2_{\pmb{ \zeta^2_k} }$ and $\tau_{\pmb{ \zeta^3_k} }=1/\sigma^2_{\pmb{ \zeta^3_k} }$. 
Our implementation of the mortality model from Equation \ref{eq:mortality_rates_equation} with interaction terms of Typ IV, can be found on Github.\footnote{\url{https://github.com/CorinnaPerchtold/Mortality_Rates}}

%The spatial Leroux random effect $\phi_i$ is not directly available in R-INLA, but can be easily built using the \textit{generic1} model, see \cite{Ugarte2014} for some code. Since we did not have information about the strength of the spatial dependence, we have chosen a non-informative prior for the spatial component $\lambda_{\pmb{\phi}}$, i.e., $\text{logit}(\lambda_{\pmb{\phi}}) \sim \text{logitbeta}(1, 1)$ and for the precision parameter $\tau_{\pmb{\phi}}=1/\sigma^2_{\pmb{\phi}}$ we chose $\log(\tau_{\pmb{\phi}}) \sim \text{logGamma}(1, 0.01)$ as suggested by \cite{Goicoa2016}.
\subsubsection{Priors and constraints}
For the intercept $\alpha_k$ we use a Gaussian prior with mean and precision equal to zero and  for the coefficients $\gamma_{1,k},\ldots, \gamma_{13,k}$ we also have mean zero and precision equal to $0.001$. Both prior assumptions are default.
Since we did not have information about the strength of the spatial dependence in the model, we have chosen a non-informative prior for the spatial component of the Leroux  effect $\lambda_{\pmb{\phi_k}}$, i.e., $\text{logit}(\lambda_{\pmb{\phi_k}}) \sim \text{logitbeta}(1, 1)$ and for the precision parameter $\tau_{\pmb{\phi_k}}$ we chose $\log(\tau_{\pmb{\phi_k}}) \sim \text{logGamma}(1, 0.01)$ as suggested by \cite{Ugarte2014}. We also imposed a sum-to-zero constraint on the spatial random effect and the RW($1$) effects. 
For the rest of the precision parameters, minimally informative priors (default) were used, i.e., $\log(\tau_{\ldots})\sim \text{ logGamma}(1,0.00005$).
Because the structure matrices of the RW($1$) effects  $R_{\pmb{\delta_k}}$ and $R_{\pmb{\psi_k}}$ are not of full rank, identifiability of the space-time, space-age and age-time effects can only be ensured by computing the null spaces of these matrices and using the obtained eigenvectors as linear constraints. The number of linear constraints is always equal to the rank deficiency of $R_{\pmb{\delta_k}}$ and $R_{\pmb{\psi_k}}$, see \cite{Schroedle2011}. To avoid confounding with the main effects, we implemented sum-to-zero constraints on each of these interaction terms. A comprehensive summary of identifiability constraints in disease mapping, can be found in \cite{Goicoa2018}.

%When working with area level data, the spatial dependency is taken into account through the neighborhood structure, see Figure \ref{fig:neighborhod-structure} for the case of Austria. Simplifying the notation introduced in the previous section so that the locations $(s_1,…, s_n)$ become $(1,…, n)$, then typically given the area $i$, its neighbors $N(i)$ are defined as the areas which share borders with it, \cite{Blangiardo2015}. The $9$ columns and rows represent the different states of Austria. In particular, the squares identify neighbouring states.

%\begin{figure}[h]
%    \centering
%    \includegraphics[scale=0.5]{adjacency matrix.png}
%    \caption{Adjacency matrix for Austria: rows and columns identify areas; squares identify neighbors.}
%    \label{fig:neighborhod-structure}
%\end{figure}
\subsubsection{Inference with INLA}

In recent years a new approach, relying on integrated nested Laplace approximations (INLA), for latent Gaussian models was introduced by Rue and Held \cite{Rue2009}.
This method is particularly effective when the latent Gaussian field is a Gaussian Markov random field with sparse precision matrix and few hyperparameters.
Rather than relying on sampling, this method approximates the marginal posterior distributions -- first for the hyperparameters, and then for the latent variables conditional on those hyperparameters -- providing efficient and accurate inference. Our model was fitted using INLA, through the corresponding R package \cite{RTeam2011}, R-INLA. This can be downloaded from the web page where information, examples, and a user's forum are available.\footnote{\url{http://www.r-inla.org/}}

\section{Discussion of Results}

The results from the Bayesian hierarchical spatio-temporal model reveal distinct effects of temperature, humidity, and temporal factors on mortality in Austria, with notable differences between males and females. The mean and median posterior draws of the estimated fixed effects are reported in Tables \ref{tab:posterior_female} and \ref{tab:posterior_male} for female and male mortality rates, respectively. 

\begin{itemize}
    \item Temperature effects
\end{itemize}
Higher maximum temperatures are significantly associated with increased mortality in both groups, but the effect size appears slightly larger for the female group compared to the male group\footnote{It is worth noting that the minimum and maximum current temperatures were standardised, such that the coefficients can directly be interpreted as effect sizes. Moreover, recall that the log-link was used, such that the coefficients need to be interpreted as multiplicative effects on the log-mortalities. To back-transform the (multiplicative) effects into the original scale, the estimated coefficients need to be exponentiated.}. More specifically, a one-standard-deviation increase in maximum temperature is associated with a 3.4\% increase in the expected mortality rate for women, and a 2.7\% increase for men. In contrast, cold stress also leads to excess mortality, particularly in super cold weeks, with men (3.0\% increase) and women (6.7\% increase) both showing a significant increase in mortality risk. This aligns with epidemiological findings that cold exposure often leads to cardiovascular and respiratory complications, particularly in elderly and vulnerable populations \citep[see, e.g.,][]{fan2023systematic,ni2024short}.

\begin{itemize}
    \item Humidity effects
\end{itemize}
Humidity-related risks show a stronger impact on female mortality, particularly under severe malaise humidity conditions (7.6\% increase) compared to men (2.1\% increase). Women may be physiologically more affected by extreme humidity, potentially due to differences in thermoregulation, hydration status, or cardiovascular responses \citep[see, e.g.,][]{armstrong2019role,chen2021role}. Interestingly, ``serious risk humidity'' shows a larger but more uncertain effect in men than in women, suggesting possible differences in behavioural responses, pre-existing health conditions, or occupational exposures. Moreover, ``strong discomfort humidity'' has a significant, positive effect for women, while it was not statistically significant for men.

One striking contrast is the effect of dry conditions in the previous week, which is significantly associated with lower mortality risk in men (15.1\% decrease), while no significant effect is observed in women. This could indicate differential adaptation mechanisms, where men may benefit more from drier air conditions, potentially reducing respiratory infections or cardiovascular strain.

% \textcolor{blue}{MISSING: female mortality significant covariates: mild week and strong discomfort humidity}

\begin{itemize}
    \item Other effects
\end{itemize}
Moreover, a higher elevation reduces mortality risk significantly in both sexes, with a stronger protective effect for men (13.5\% per 1000m) than for women (11.3\% decrease per 1000m). This may suggest that men living at higher altitudes adapt better to lower oxygen levels, or that there are indirect lifestyle and socioeconomic factors contributing to the observed difference.

To assess the goodness-of-fit of our models, we considered the Deviance Information Criterion (DIC). The saturated model DIC values suggest that both models significantly outperform their saturated counterparts, indicating that the hierarchical structure and interaction effects contribute meaningfully explain the mortality rates. The slightly higher complexity of the female model (larger effective parameter count) may reflect stronger interaction effects or higher variability in female mortality patterns in response to temperature and humidity stressors. Overall, the DIC values confirm the models’ appropriateness in capturing the spatio-temporal mortality dynamics.

\begin{table}[ht]
    \centering
    \caption{Posterior estimates for fixed effects in the Bayesian hierarchical spatio-temporal model (Female mortality). Statistically significant effects are highlighted in bold.}
    \label{tab:posterior_female}
    \begin{tabular}{lccc}
        \hline
        \textbf{Variable} & \textbf{Mean} & \textbf{Median} & \textbf{95\% CI} \\
        \hline
        \multicolumn{4}{l}{\emph{Temperature Effects}} \\
        $\quad$ scale(Temp min mean) & -0.001 & -0.001 & (-0.016, 0.014) \\
        $\quad$ {scale(Temp max mean)} & \textbf{0.034} & \textbf{0.034} & {(0.018, 0.050)} \\
        $\quad$ Hot week & -0.121 & -0.121 & (-0.386, 0.144) \\
        $\quad$ {Cold week} & \textbf{0.047} & \textbf{0.047} & {(0.031, 0.063)} \\
        $\quad$ {Super cold week} & \textbf{0.067} & \textbf{0.067} & {(0.045, 0.089)} \\
        $\quad$ Mild week & \textbf{0.023} & \textbf{0.023} & (0.012, 0.035) \\[.2cm]
        \multicolumn{4}{l}{\emph{Humidity Effects}} \\
        $\quad$ scale(Humidity mean) & 0.000 & 0.000 & (-0.005, 0.006) \\
        $\quad$ Last week was dry & 0.018 & 0.018 & (-0.110, 0.146) \\
        $\quad$ {Increased risk humidity} & \textbf{0.089} & \textbf{0.089} & {(0.038, 0.140)} \\
        $\quad$ Serious risk humidity & 0.328 & 0.328 & (-0.039, 0.695) \\
        $\quad$ Strong discomfort humidity & \textbf{0.015} & \textbf{0.015} & (0.003, 0.028) \\
        $\quad$ {Severe malaise humidity} & \textbf{0.076} & \textbf{0.076} & {(0.058, 0.097)} \\[.2cm]
        \multicolumn{4}{l}{\emph{Other Factors}} \\
        $\quad$ {Elevation} & \textbf{-0.113} & \textbf{-0.113} & {(-0.157, -0.070)} \\[.2cm]
        \multicolumn{4}{l}{\emph{Goodness-of-fit}} \\
        $\quad$ DIC                   & 999046.61  &   & \\
        $\quad$ DIC (saturated model) & 376571.91  &   &   \\
        $\quad$ Effective number of parameters & 966.09  &   &   \\
        \hline
    \end{tabular}
\end{table}

\begin{figure}
    \includegraphics[width=0.5\textwidth]{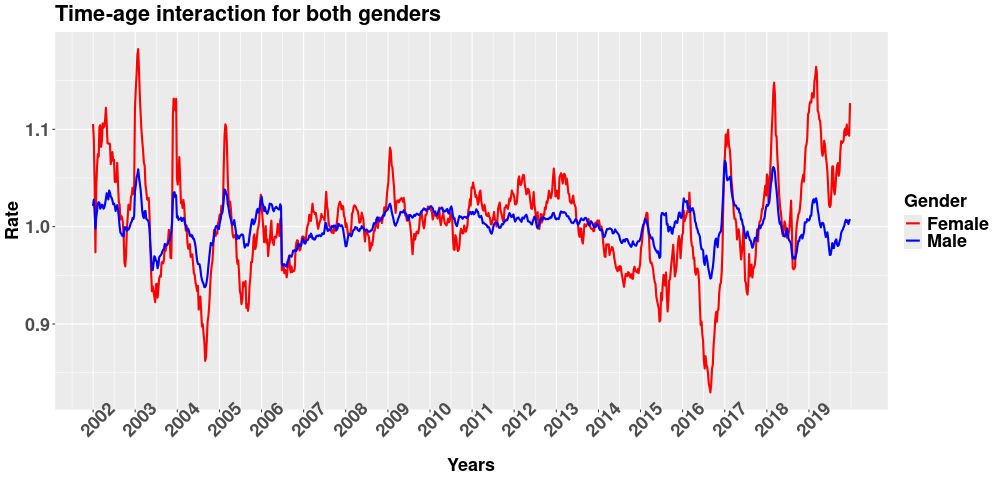}
    \includegraphics[width=0.5\textwidth]{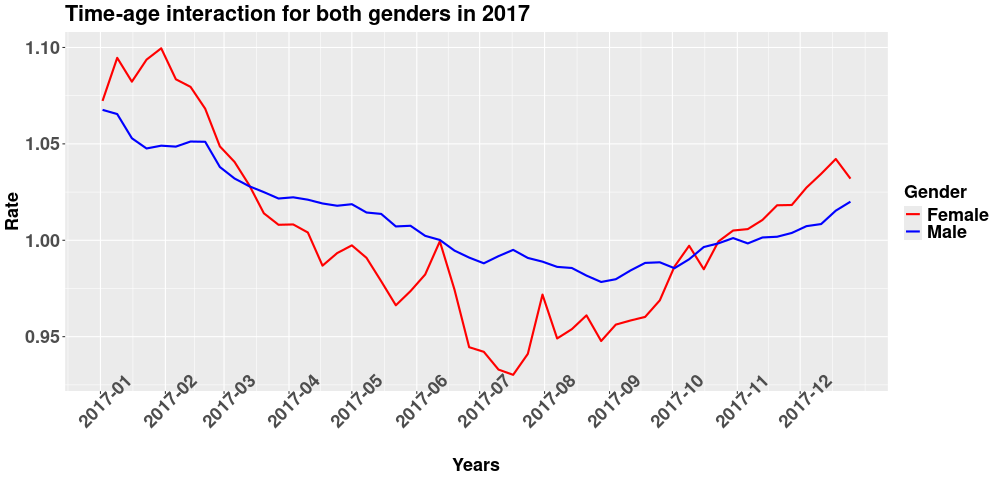}
    \caption{Time-age interaction effect for both genders for the entire time horizon averaged across all age groups (left) and for 2017 for all age groups (right).}
    \label{fig:Time-Age}
\end{figure}

\begin{table}[ht]
    \centering
    \caption{Posterior estimates for fixed effects in the Bayesian hierarchical spatio-temporal model (Male mortality). Statistically significant effects are highlighted in bold.}  \label{tab:posterior_male}
\begin{tabular}{lccc}
\hline
 \textbf{Variable} & \textbf{Mean} & \textbf{Median} & \textbf{95\% CI} \\
\hline
\multicolumn{4}{l}{\emph{Temperature Effects}} \\
 $\quad$ scale(Temp min mean) & -0.011 & -0.011 & (-0.027, 0.005) \\
 $\quad$ {scale(Temp max mean)} & \textbf{0.027} & \textbf{0.027} & {(0.011, 0.043)} \\
$\quad$ Hot week & -0.125 & -0.125 & (-0.407, 0.158) \\
$\quad$ {Cold week} & \textbf{0.021} & \textbf{0.021} & {(0.005, 0.038)} \\
$\quad$ {Super cold week} & \textbf{0.030} & \textbf{0.030} & {(0.008, 0.052)} \\
$\quad$ Mild week & 0.007 & 0.007 & (-0.004, 0.019) \\[.2cm]
\multicolumn{4}{l}{\emph{Humidity Effects}} \\
$\quad$ scale(Humidity mean) & 0.000 & 0.000 & (-0.006, 0.005) \\
$\quad$ {Last week was dry} & \textbf{-0.151} & \textbf{-0.151} & {(-0.290, -0.012)} \\
$\quad$ Increased risk humidity & -0.007 & -0.007 & (-0.061, 0.047) \\
$\quad$ Serious risk humidity & 0.070 & 0.070 & (-0.343, 0.482) \\
$\quad$ Strong discomfort humidity & -0.006 & -0.006 & (-0.019, 0.006) \\
$\quad$ {Severe malaise humidity} & \textbf{0.021} & \textbf{0.021} & {(0.000, 0.042)} \\[.2cm]
\multicolumn{4}{l}{\emph{Other Factors}} \\
$\quad$ {Elevation} & \textbf{-0.135} & \textbf{-0.135} & {(-0.181, -0.089)} \\[.2cm]
 \multicolumn{4}{l}{\emph{Goodness-of-fit}} \\
$\quad$ DIC                   & 1038526.69 &   & \\
$\quad$ DIC (saturated model) & 394003.53  &   &   \\
$\quad$ Effective number of parameters & 795.28  &   &   \\
        \hline
    \end{tabular}
\end{table}
\vspace{10mm}
\begin{itemize}
    \item Random effects
\end{itemize}
Below, we turn our focus on the estimated time-age, space-time and space-age interaction random effects. The time-age interactions are depicted in Figure \ref{fig:Time-Age}. Overall, the random effects for the female subgroup tend to be higher in absolute terms (i.e., absolute deviations from one), meaning that age-specific mortality deviations from expected trends were more pronounced for women compared to men. This suggests that women’s age-related mortality risks are more sensitive to temporal fluctuations. Over the entire analysed time period, we observe periods of high variation in random effects and periods of lower variations. For instance, the 2017 influenza season stands out as a critical period, displaying the expected epidemiological pattern of peak mortality in January–February, a decrease in summer, and a renewed increase toward the end of the year. The nearly identical impact across age groups suggests that the 2017 influenza epidemic was particularly severe and widespread (about 440000 influenza incidences), affecting all vulnerable populations similarly \citep[see also][]{redlberger2020heterogeneity}. In contrast, periods of low variation in random effects, such as between 2008 to 2014, suggest that mortality patterns were more stable, possibly due to milder seasonal influenza outbreaks or the absence of extreme climatic conditions.

\begin{figure}
    \includegraphics[width=0.5\linewidth]{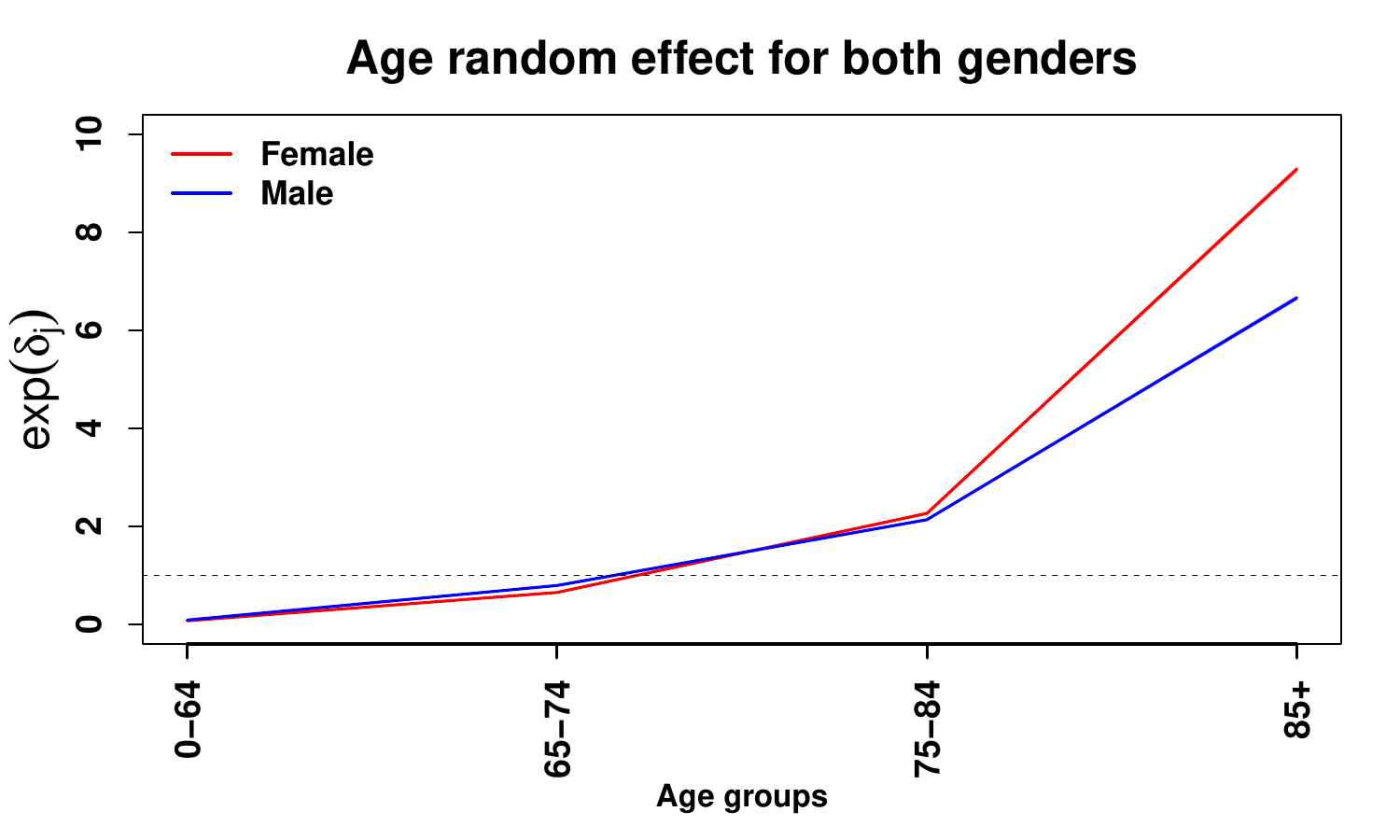}
        \includegraphics[width=0.5\linewidth]{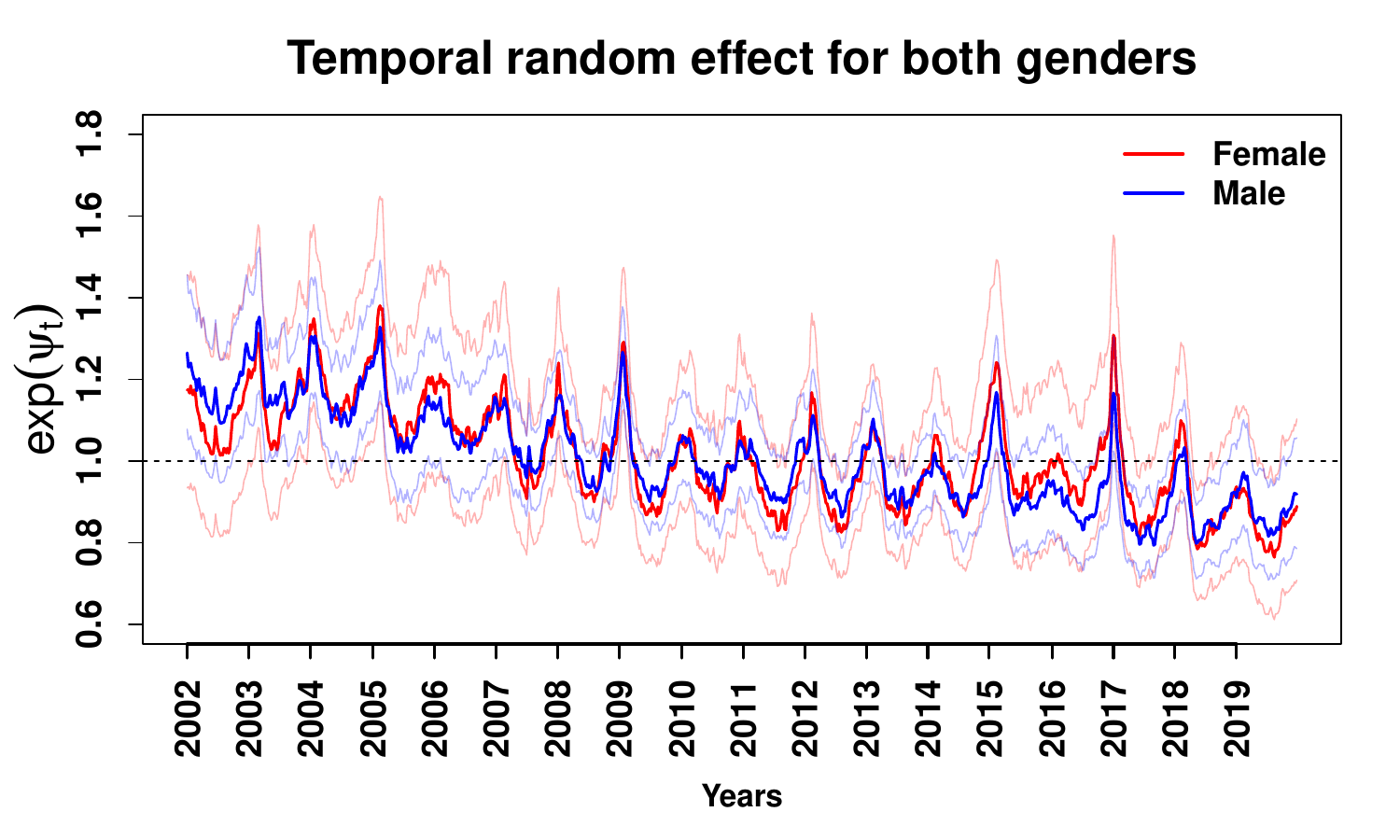}
    \caption{Random age effect and temporal effect for both genders.}
    \label{fig:age_random_effect}
\end{figure}

\begin{figure}
    \centering
    \includegraphics[width=0.9\linewidth]{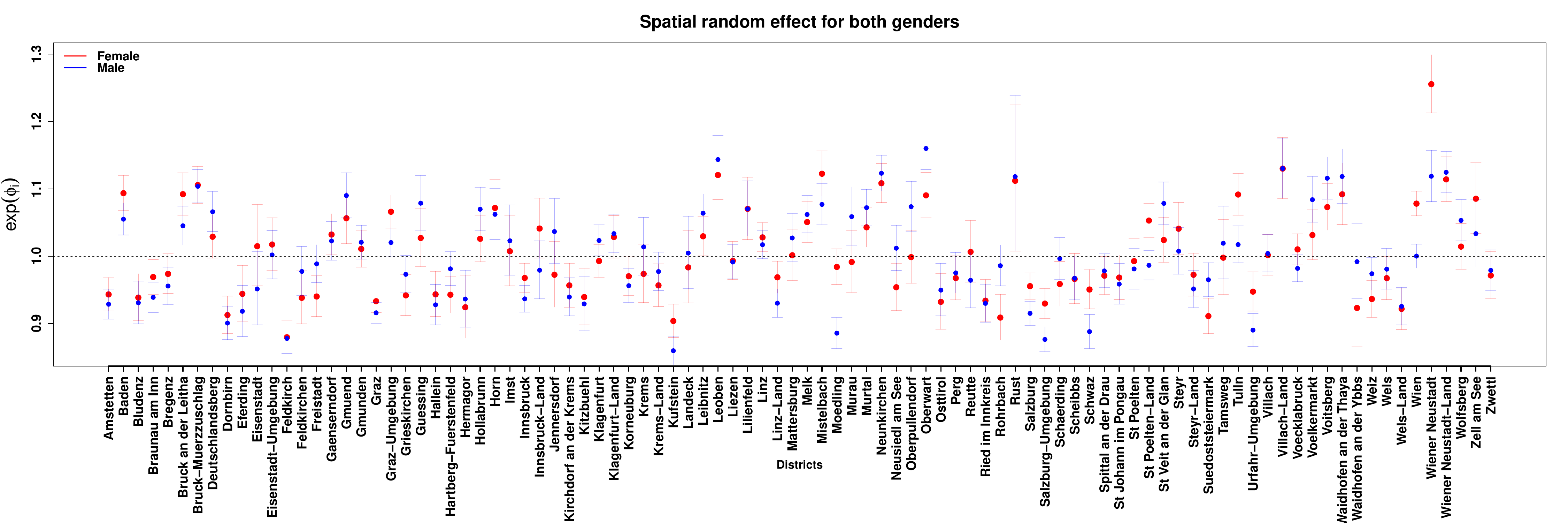}
    \caption{Random spatial effect for both genders.}
    \label{fig:space_random_effect}
\end{figure}

Figures \ref{fig:age_random_effect} and \ref{fig:space_random_effect} display the age-group random effects (\ref{fig:age_random_effect}, left), the temporal random effects (\ref{fig:age_random_effect}, right), and the spatial random effect, respectively. The age random effects show a steep increase in mortality risk with advancing age, particularly among individuals aged 85 and older, reflecting the well-established vulnerability of the elderly population. Notably, the gender gap widens in the oldest age group, with higher relative mortality effects for women, suggesting that while women tend to live longer, they may experience an increased mortality intensity in the oldest age group (85+). The temporal random effects reveal a clear seasonal pattern in mortality for both genders, with winter peaks and summer troughs, and slightly higher variability in female mortality over time. The spatial random effects indicate largely similar regional mortality patterns for men and women, with only a few districts (e.g., Vienna) exhibiting notable gender-specific deviations.
% \textcolor{red}{Corinna: Kannst du Grafiken/Tabellen für phi, delta und psi erstellen? } \textcolor{blue}{Philipp:ja, siehe Figure \textbf{NOTE HERE:} only one age line displayed since the others are the same, see Figure \ref{fig:Time-Age}.}

%\begin{figure}
%    \includegraphics[width=0.5\linewidth]{time_random_effect.png}
%    \caption{Random time effect for females and males.}
%    \label{fig:time_random_effect}
%\end{figure}

\begin{figure}
    \centering
     \includegraphics[width=\textwidth, trim=0 35mm 0 35mm,clip]{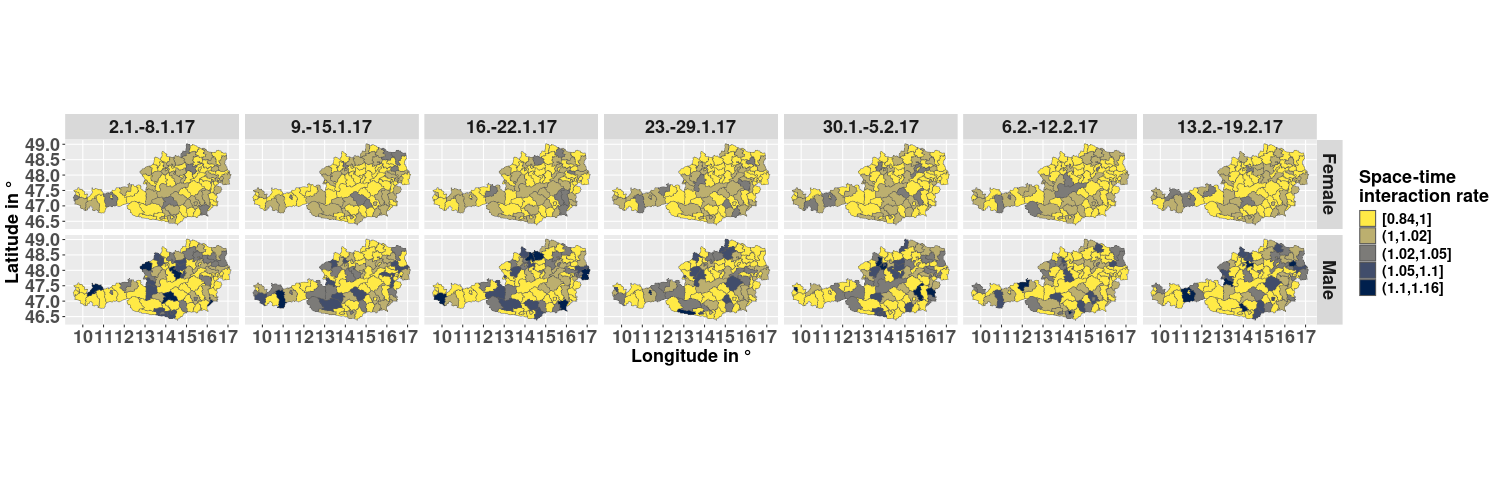}
    \caption{Space-time interaction effect for both genders from $2.1-19.2.2017$.} 
    \label{fig:Space-Time}
\end{figure}

The space-time interaction effects displayed in Figure \ref{fig:Space-Time} reveal how mortality trends evolve differently across regions over time, beyond what is explained by overall temporal or spatial trends alone. These effects capture localised deviations in mortality risk, influenced by regional climatic conditions, demographic composition, healthcare infrastructure, and environmental stressors.

There are periods of heightened variability in the space-time interactions, which typically coincide with extreme climate events, severe influenza seasons, or localised events. For instance, during the 2017 influenza season, the space-time effects indicate a strong winter peak in January–February, mainly in the region of Tyrol, 
% \textcolor{red}{Corinna: Zeitpunkt 791, korrekt? Ist die Farbskala in allen Karten gleich oder unterschiedlich?}
%\textcolor{blue}{Start 2017 mit Zeitpunk 784 u die Farbskala ist gleich }
a decrease in summer, and a resurgence in late autumn—a seasonal pattern seen across regions. However, the magnitude of these effects varies geographically, likely reflecting differences in regional healthcare capacity, vaccination rates, or demographic vulnerability. Regions with older populations or limited healthcare accessibility may have exhibited stronger deviations from the expected mortality trend. Conversely, periods of lower variation in space-time interactions indicate a more uniform mortality trend across regions. In these periods, spatial disparities in mortality risk were less pronounced, suggesting that climatic and epidemiological factors affected regions more uniformly.

% \subsection{Space-Time interaction effect}
 
\begin{figure}
    \centering
\includegraphics[width=\textwidth, trim=0 30mm 0 0,clip]{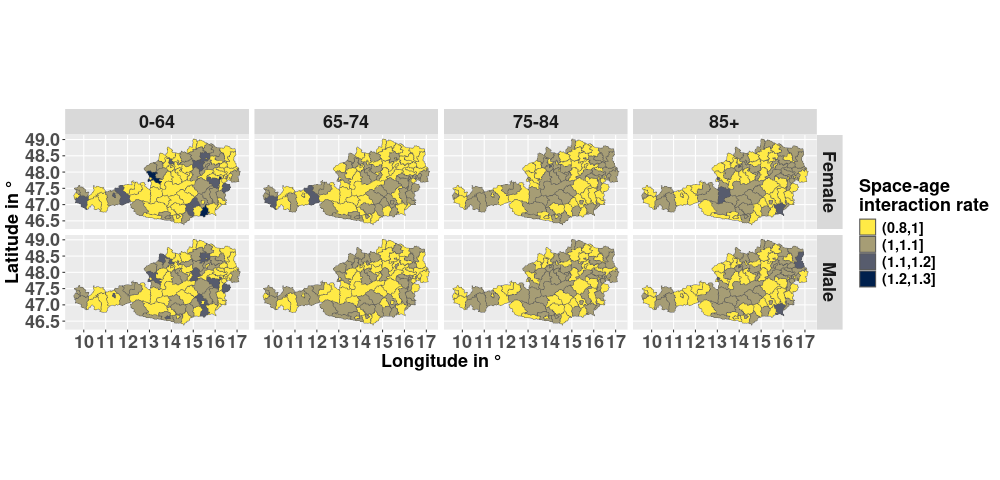}
    \caption{Space-age interaction effect for both genders and four age groups.}
    \label{fig:Space-Age}
\end{figure}

Figure \ref{fig:Space-Age} shows the space-age interaction random effects, revealing that age-specific mortality patterns vary across different regions, beyond what is explained by overall spatial or age-related trends. This suggests that factors such as healthcare access, socioeconomic status, and environmental exposures may differentially impact female mortality across regions. In contrast, male mortality patterns appear more geographically uniform, potentially due to higher baseline mortality rates associated with causes like cardiovascular diseases and accidents, which are prevalent across various regions. These findings align with existing literature indicating that women generally have a longer life expectancy than men globally, though the complexity of this advantage challenges simplistic explanations in Zarulli and Salinari \citep{zarulli2024gender}.  Understanding these gender-specific spatial variations is crucial for developing targeted public health interventions that address the unique needs of each demographic group. 
% \subsection{Space-Time interaction effect}

%\subsection{Space-Age interaction effect}
 %\textbf{NOTE HERE:} did not find out yet what missing values are meant since all values are captured within the legends intervals

% \subsection{Time-Age interaction effect}

%\subsection{Convergence?}
%\begin{figure}
%    \centering
%    \includegraphics[width=0.5\linewidth]{image.png}
%    \caption{just so I dont forget}
%    \label{fig:enter-label}
%\end{figure}

\section{Conclusion}
In this study, we presented a novel gender-specific extension of a spatio-temporal hierarchical model for mortality in Austria, incorporating 13 fixed effects and random as well as interaction effects for space, time, and age. This framework enables a detailed decomposition of mortality dynamics at the district level and allows us to quantify the impact of climatic and environmental stressors in a gender-sensitive manner using harmonised, high-resolution data from multiple sources.

Our findings confirm that temperature extremes -- both heat and cold -- are associated with elevated mortality risks in both genders. While higher maximum temperatures lead to increased mortality in men and women, the effect is more pronounced in women. Cold stress, particularly during extremely cold weeks, results in a sharper increase in mortality for women ($6.7\%$) than for men ($3.0\%$). Humidity also shows a gendered impact: severe malaise humidity conditions are associated with a $7.6\%$ increase in female mortality, compared to only $2.1\%$ in men. Additionally, dry conditions in the preceding week appear protective for men but not for women, suggesting possible physiological or behavioural differences in adaptation.

We also observe that elevation has a protective effect for both sexes, though stronger in men. This may point to gendered adaptation to environmental or lifestyle factors associated with altitude. Beyond fixed effects, our random effects structure highlights how mortality patterns fluctuate across districts, age groups and over time. Notably, female age-specific mortality shows greater temporal sensitivity, with heightened variation during, e.g., the 2017 influenza season. Space-time and space-age interaction effects further underscore how regional characteristics interact with age and time in shaping mortality risk.

Overall, these results demonstrate the utility of our extended model in capturing gender- and region-specific mortality responses to environmental stressors. The approach can serve as a basis for more targeted public health interventions and improved understanding of demographic vulnerabilities under changing climatic conditions.

Building on these results, future research could explore several extensions and refinements. First, integrating specific disease burdens (e.g., respiratory or cardiovascular outcomes) into the modelling framework could help better understand underlying health mechanisms. Second, the lagged effects of heat -potentially extending over several days or weeks-warrant more detailed investigation. Third, alternating effect between heat and cold stress could be more explicitly modelled. Finally, future work could benefit from automated covariate selection techniques to streamline model complexity and improve interpretability.

% Furthermore, our findings provide a foundation for future research on climate-sensitive health risks. Potential directions include examining the effects of deprivation and climate change on mortality rates, as explored by Yiu et al.\ \cite{Torsten} with no climate change as a variable, or advancing personalised treatment strategies within survival models like in Efthimiou et al.\ \cite{treatmentchoices} and broader healthcare contexts like in Gutzeit et al.\ \cite{healthcare}.

\section{Data Availability}
The meteorological data was downloaded from the data hub of GeoSphere, see \url{https://data.hub.geosphere.at/dataset/klima-v2-1d}. The elevation map is taken from the Global Administrative Areas and can be found at \url{https://gadm.org/download_country.html}.
The gender-specific population data for the age groups and districts over the years $2002-2019$ were downloaded from the statistical database STATcube, available here \url{https://statcube.at/statistik.at/ext/statcube/jsf/dataCatalogueExplorer.xhtml}.

	\bibliographystyle{plain}
		\bibliography{lib_tex.bib}

\end{document}